\documentclass[10pt,conference]{IEEEtran}
\IEEEoverridecommandlockouts
\usepackage{balance} 
\usepackage{cite}
\usepackage{amsmath,amssymb,amsfonts}
\usepackage{algorithmic}
\usepackage{graphicx}
\usepackage{textcomp}
\usepackage{xcolor}
\usepackage{pifont}
\usepackage{subfigure}
\usepackage{multirow}
\usepackage{bm}
\usepackage{multicol}
\usepackage{threeparttable}
\usepackage[ruled,linesnumbered]{algorithm2e}
\usepackage{booktabs}
\usepackage{fontawesome5} 
\usepackage{colortbl}
\usepackage{color}
\usepackage{multibib}
\usepackage{tcolorbox}
\usepackage{hyperref}
\usepackage{varwidth}
\usepackage{lettrine}

\usepackage{hyperref}
\hypersetup{hidelinks,
	colorlinks=true,
	allcolors=black,
	pdfstartview=Fit,
	breaklinks=true}

\tcolorboxenvironment{greyref}{arc=1mm, boxrule =0mm, leftrule = 0.5mm}

\def\BibTeX{{\rm B\kern-.05em{\sc i\kern-.025em b}\kern-.08em
    T\kern-.1667em\lower.7ex\hbox{E}\kern-.125emX}}
\begin{document}

\newcommand{\hy}[1]{\mytodored{[HY: #1]}}
\newcommand{\mytodored}[1]{\textcolor{red}{~{\sf}~#1}}

\newcommand{\lhy}[1]{\mytodoblue{[LHY: #1]}}
\newcommand{\mytodoblue}[1]{\textcolor{blue}{~{\sf}~#1}}

\title{Revisiting and Improving Retrieval-Augmented Deep Assertion Generation
}

\makeatletter
\newcommand{\linebreakand}{%
  \end{@IEEEauthorhalign}
  \hfill\mbox{}\par
  \mbox{}\hfill\begin{@IEEEauthorhalign}
}
\makeatother

\author{
  \IEEEauthorblockN{
  Weifeng Sun\IEEEauthorrefmark{1},
  Hongyan Li\IEEEauthorrefmark{1}\thanks{\IEEEauthorrefmark{1} Both authors contributed equally to this research},
  Meng Yan\IEEEauthorrefmark{2}\thanks{\IEEEauthorrefmark{2} Meng Yan is the corresponding author},
  Yan Lei,
  Hongyu Zhang
  }
  
  \IEEEauthorblockA{ School of Big Data and Software Engineering, Chongqing University, Chongqing, China \\
                    \{weifeng.sun, hongyan.li, mengy, yanlei, hyzhang\}@cqu.edu.cn}
}

\maketitle

\begin{abstract}
Unit testing 
validates
the correctness of the unit under test and has become an essential activity in software development 
process.
A unit test consists of a test prefix that drives the unit under test into a particular state, and a test oracle (e.g., assertion), which specifies the behavior in that state.
To reduce manual efforts in
conducting unit testing, Yu et al. proposed an integrated approach (\textit{integration} for short), combining information retrieval with a deep learning-based approach, to generate assertions for a unit test. 
Despite being promising, there is still a knowledge gap as to why or where \textit{integration} works or does not work.
In this paper, we describe an in-depth analysis of the effectiveness of \textit{integration}. 
Our analysis shows that:
\ding{172} The overall performance of \textit{integration} is mainly due to its success in retrieving assertions.  
\ding{173} 
\textit{integration} struggles to understand the semantic differences between the retrieved focal-test (\textit{focal-test} includes a test prefix and a unit under
test) and the input focal-test, resulting in many tokens being incorrectly modified;
\ding{174} \textit{integration} is limited to specific types of edit operations (i.e., replacement) and cannot handle token addition or deletion.
To improve the effectiveness of assertion generation, this paper proposes a novel retrieve-and-edit approach named \textsc{EditAS}. 
Specifically, \textsc{EditAS} first retrieves a similar focal-test from a pre-defined corpus and treats its assertion as a prototype. 
Then, \textsc{EditAS} reuses the information in the prototype and edits the prototype automatically.
\textsc{EditAS} is more generalizable than \textit{integration} because it can 
\ding{182} comprehensively understand the semantic differences between input and similar focal-tests; 
\ding{183} apply appropriate assertion edit patterns with greater flexibility; and 
\ding{184} generate more diverse edit actions than just replacement operations.
We conduct experiments on two large-scale datasets and the experimental results demonstrate that \textsc{EditAS} outperforms the state-of-the-art approaches, with an average improvement of 10.00\%-87.48\% and 3.30\%-42.65\% in accuracy and BLEU score, respectively.

\end{abstract}


\section{Introduction}

Unit testing is a crucial activity of software development, which involves testing an individual unit of software applications, such as a method, a class, or a module.
While integration and system testing assess the overall performance of a system, unit testing focuses on validating that each unit of code works as intended and conceived by the developer, thereby detecting and diagnosing failures before they propagate throughout the system and preventing regressions.
Effective unit tests can improve the quality of software, reduce the incidence, and cost of software failures~\cite{Hartman2002, Planning2002}, as well as enhance the entire software development process.

\begin{figure}[t]
	\graphicspath{{graphs/}}
	\centering
	\includegraphics[width=\linewidth]{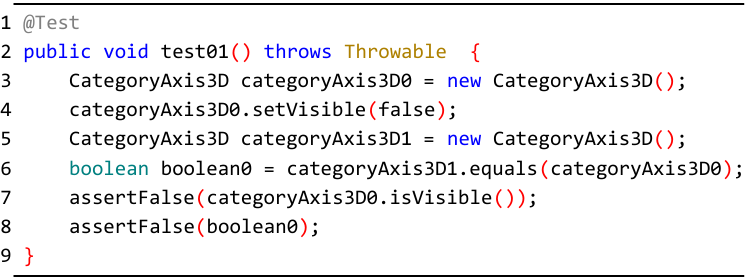}
	\caption{Example of a unit test case.}
	\label{FIG:example}
\end{figure}

A unit test comprises a \emph{test prefix}, which is a series of statements that manipulate a unit under test to attain a specific state, and a \emph{test oracle}, which typically includes an assertion that specifies the expected behavior under that state~\cite{Dinella2022}.
For instance, in the unit test illustrated in Figure~\ref{FIG:example}, lines 3-6 constitute the test prefix, which creates two \texttt{CategoryAxis3D} objects, with \texttt{categoryAxis3D0} set to invisible; testers utilize a boolean variable to check whether \texttt{categoryAxis3D0} and \texttt{categoryAxis3D1} are equal. 
On lines 7-8, the assertion specifies the expected outcome, which tests that after executing the test prefix, the visibility property of \texttt{categoryAxis3D0} should be \texttt{False} and that the two objects should not be equivalent.

Despite the significant benefits of testing, creating effective unit tests is a non-trivial and time-consuming task. 
Previous studies have indicated that developers can spend more than 15\% of their time on test generation~\cite{Daka2014}. 
To streamline unit test generation, various automated testing tools have been proposed, such as Randoop~\cite{Pacheco2007a} and EvoSuite~\cite{Fraser2011}.
However, these test-generation tools prioritize generating high-coverage tests over meaningful assertions and 
face challenges in comprehending the intended program behavior.
For example, an industrial evaluation~\cite{Almasi2017} of assertions generated by EvoSuite has shown that ``in manually written tests, the assertions are meaningful and useful unlike the generated ones.'' 
As a result, a lot of manual effort is still required in conducting unit testing.


To reduce the effort for oracle generation, Watson et al. introduced ATLAS~\cite{Watson2020}, a deep learning (DL)-based technique that trains a neural generative model on an extensive corpus of existing unit tests.
ATLAS operates by taking a pair consisting of a test prefix and its \emph{focal method} (i.e., the method under test) that encompasses both the method names and the method bodies.
For consistency with prior research~\cite{Yu2022}, we refer to such a pair as \emph{focal-test}.
ATLAS then generates an assertion for the focal-test from scratch.
Neural models, unlike specification-mining-based approaches, are more flexible, particularly when documentation is imprecise or incomplete.
Thus, ATLAS offers a more adaptable solution to the test assertion problem.
However, the effectiveness of ATLAS is limited by several issues:
1) ATLAS generally prefers high-frequency words in the corpus and may have trouble with low-frequency words, such as project-specific identifiers.
2) ATLAS has poor performance when generating long sequences of tokens as assertions.

Recently, Yu et al.~\cite{Yu2022} proposed an information retrieval (IR)-based assertion generation method, including IR-based assertion retrieval ($IR_{ar}$) and retrieved-assertion adaptation ($RA_{adapt}$) techniques. 
$IR_{ar}$ takes the same input as ATLAS and retrieves the most similar assertion to the given focal-test based on the Jaccard similarity coefficient~\cite{Tanimoto1958}.
Then, $RA_{adapt}$ further adjusts the tokens in the retrieved assertion based on the context.
Furthermore, an integrated approach~\cite{Yu2022} (abbreviated as \textit{Integration}) that combines the IR-based approach with a DL-based approach has been proposed to improve assertion generation capabilities.
The integrated method verifies the compatibility between the retrieved assertion and the current focal-test. If the compatibility exceeds a threshold, the retrieved assertion is returned as the final result. Otherwise, the DL-based method generates the assertion.
Experimental results show that the integrated approach achieves higher accuracy and BLEU score than ATLAS. While the performance is a notable achievement in assertion generation research, knowledge gaps still exist regarding why or where the proposed technique is effective or ineffective.


To fill this gap in the literature, we first conduct a comprehensive evaluation of \textit{Integration} to gain a better understanding of its application scenarios.
We conduct the empirical assessment on two public datasets adopted by Yu et al.~\cite{Yu2022}, namely $Data_{old}$ and $Data_{new}$. 
The main findings are:

\begin{itemize}
    \item \textit{Integration} relies mainly on the IR-based method for generating assertions. 
    For instance, \textit{Integration} utilizes the IR-based approach to generate 80.06\% of the assertions for $Data_{new}$.
    Additionally, for $Data_{old}$ and $Data_{new}$, 83.38\% and 92.47\% of correct assertions generated by the IR-based approach are identical to the retrieved assertions.
    
    \item \textit{Integration} only replaces tokens during its adaptation operation, making it challenging to generalize to complex assertion generation scenarios.

    \item \textit{Integration} incorrectly modifies assertions frequently, even when the retrieved assertion matches the ground truth exactly.
    When one token needs to be modified to get the correct result, the average accuracy of \textit{Integration} is just 20.14\% and drops to only 1.91\% for more than five tokens.

    \item \textit{Integration} fails to comprehend the semantic differences between focal-tests, resulting in many cases where retrieved assertions require modifications, but it is returned directly as expected assertion.

\end{itemize}

Overall, our empirical study reaffirms previous findings that based on the similarity of focal-test, we can always find ``almost correct" assertions that are very similar to the correct ones. 
However, we also identify several limitations that restrict the effectiveness and generalizability of \textit{Integration}.
On the one hand, the single token replacement operation struggles with complex assertion edit scenarios, resulting in low accuracy (1.9\%) of \textit{Integration} when modifying more than five tokens.
On the other hand, \textit{Integration}'s inability to comprehend semantic differences between input and similar focal-tests causes it to make incorrect editing actions or fail to edit retrieved assertions when necessary.

To alleviate the above-mentioned limitations and achieve higher levels of performance, this paper proposes \textsc{EditAS}, a novel retrieve-and-edit approach for assertion generation. 
The effectiveness of IR implies that similar focal-tests' assertions can be reused.
In other words, certain tokens in the expected assertion are also highly probable to appear in the retrieved assertion.
The improvements by the specification-mining-based and retrieved-assertion adaptation approaches have revealed the significance of assertion patterns.
Inspired by this, \textsc{EditAS} views the assertion from a similar focal-test as a prototype and leverages a neural sequence-to-sequence model to learn the assertion edit patterns used to modify the prototype.
Our motivation is that the retrieved assertion guides the neural model on ``how to assert'' and the assertion edit pattern highlights to the neural model ``what to assert".

\textsc{EditAS} consists of two major components: a Retrieval component and an Edit component.
Given an input focal-test, the Retrieval component obtains its similar focal-test from a corpus and utilizes the retrieved focal-test's corresponding assertion as a prototype.
In the Edit component, a sequence-to-sequence neural network is trained to edit the prototype, based on edit sequences representing the semantic differences between the input and the similar focal-test.
On the one hand, \textsc{EditAS} effectively mitigates the long assertion generation problem, which is the performance bottleneck of ATLAS, by editing assertions instead of generating them from scratch.
Additionally, \textsc{EditAS} is more generalizable than \textit{Integration} because it can
1) comprehensively understand the semantic differences between focal-tests;
2) apply appropriate assertion edit patterns with greater flexibility; and 
3) generate more diverse edit actions than just replacement operations.

The IR-based assertion retrieval method ($IR_{ar}$), the retrieved-assertion adaptation methods including $RA_{adapt}^{H}$ and $RA_{adapt}^{NN}$, the integrated approach \textit{Integration}, and ATLAS are used as baselines.
We evaluate \textsc{EditAS} and the baselines on both $Data_{old}$ and $Data_{new}$ datasets in terms of accuracy and BLEU.
The evaluation results show that \textsc{EditAS} outperforms all baselines across all metrics in assertion generation. 
Specifically, \textsc{EditAS} achieves the highest accuracy, outperforming the baselines by 14.87\%-70.15\% and 5.12\%-104.80\% on the two datasets, respectively.

In summary, the contributions of this paper include:

\begin{enumerate}
    \item[(1)] We conduct an in-depth analysis of the state-of-the-art assertion generation method that combines DL and IR techniques. 
    Our analysis results provide valuable insights for future research in this direction.
    \item[(2)] We propose a novel retrieve-and-edit approach, namely \textsc{EditAS}, for assertion generation. 
    \textsc{EditAS} utilizes assertions from similar focal-tests as prototypes and uses a neural sequence-to-sequence model to learn the assertion edit patterns.
    \item[(3)] We conduct extensive experiments to evaluate our approach. 
    The experimental results show that \textsc{EditAS} significantly outperforms all baselines.
    \item[(4)] We open-source our replication package\footnote{https://github.com/swf-cqu/EditAS}, including the dataset, the source code of \textsc{EditAS}, the trained model, and assertion generation results, for follow-up studies.
    
\end{enumerate}

\section{Background and related work}
\label{background}

\subsection{DL-based Assertion Generation}

With the rise of deep learning (DL), an increasing number of software engineering tasks can be effectively tackled using advanced DL techniques, such as 
code search~\cite{Gu2016, Gu2018}, automated program repair~\cite{Chen2021, Hata2018, Mesbah2019, Tufano2018, Zhang2023}, fault diagnosis~\cite{Lou2021}, code summarization~\cite{Hu2018, Li2020, Hu2018a, Zhang2020}, and code clone detection~\cite{Wang2020a, Wei2017, Yu2019, Zhang2019, Zhao2018}.
Watson et al.~\cite{Watson2020} recently proposed ATLAS, the first DL-based assertion generation method.
ATLAS employs Neural Machine Translation (NMT) to generate assertions for a given \emph{focal-test}, which consists of a test method without any assertion (i.e., test prefix) and its focal method (i.e., the method under test), including both method names and method bodies. 
To develop ATLAS, Watson et al. first extracted Test-Assert Pairs (TAPs) from GitHub projects that use the JUnit testing framework. 
Each pair consists of a focal-test and its corresponding assertion.
The initial TAP set is then preprocessed into two datasets: 
1) raw source code, where TAPs are simply tokenized (see the top part of Figure~\ref{FIG:version}), and 
2) abstract code, where uncommon tokens are further represented by their respective types and sequential IDs from the raw source code (see the bottom part of Figure~\ref{FIG:version}).
Ultimately, ATLAS generates a meaningful assertion to verify the correctness of the focal method.

Recent research~\cite{Tufano2022, Mastropaolo2021, Mastropaolo2023} has explored the potential of pre-trained models, such as T5 and BART, for supporting the assertion generation task through pre-training and fine-tuning.
Specifically, these approaches involve pre-training a transform model using a large corpus of either source code or English language, followed by fine-tuning it for assertion generation.
Dinella~\cite{Dinella2022} proposed TOGA to address the assertion generation problem by using a ranking architecture over a set of candidate test assertions.
TOGA employs grammar along with type-based constraints to limit the generation space of candidates and uses a transformer-based neural approach to obtain the most probable assertions.
We find that the formal implementation of TOGA~\cite{toga, toga_is} requires a seed/approximate assertion to determine the variables for constructing and optimizing the candidate assertion set. 
Also, the seed/approximate assertion needs to be created manually or via EvoSuite tool~\cite{Fraser2011}.
Unlike TOGA, this paper focuses on the problem of assertion generation using only focal-test.

\begin{figure}[t]
	\graphicspath{{graphs/}}
	\centering
	\includegraphics[width=\linewidth]{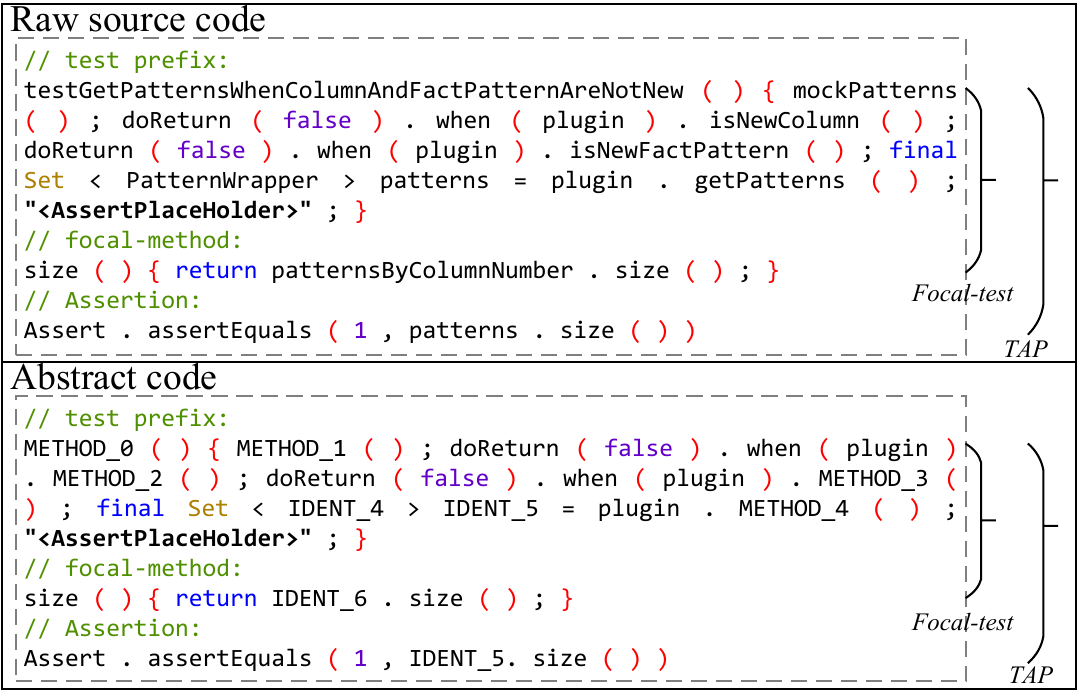}
	\caption{Abstraction process.}
	\label{FIG:version}
\end{figure}

\subsection{Combining Information Retrieval and Deep Learning for Assertion Generation}

The community has been working to boost DL techniques in software engineering tasks by leveraging Information Retrieval (IR) techniques and achieved promising results. 
Drawing on the idea of ``combining IR and DL", Yu et al.~\cite{Yu2022} proposed a novel integration approach to tackle the assertion generation problem.
Their contributions include two IR-based approaches for assertion generation, namely IR-based assertion retrieval ($IR_{ar}$) and retrieved-assertion adaptation ($RA_{adapt}$).

The basic idea of $IR_{ar}$ is to retrieve the assertion whose corresponding focal-test has the highest similarity (e.g., Jaccard~\cite{Tanimoto1958} similarity) with the given focal-test, and the retrieved assertion is returned as an expected assertion.
As $IR_{ar}$ does not always retrieve completely accurate assertions, $RA_{adapt}$ has been proposed to automatically adapt retrieved assertions to the right forms utilizing contextual information.
For a retrieved assertion, $RA_{adapt}$ performs the following adaptation process:
1) deciding whether the assertion should be modified; 
2) deciding which token (i.e., invoked method, variable, or constant) should be modified;
3) deciding what value a candidate token should be replaced with.
Yu et al. have proposed two replacement strategies for determining the replacement value: one based on heuristics ($RA_{adapt}^{H}$) and the other on neural networks ($RA_{adapt}^{NN}$).
$RA_{adapt}^{H}$ utilizes lexical similarity for code replacement.
In contrast to $RA_{adapt}^{H}$, $RA_{adapt}^{NN}$ further enhances lexical similarity by incorporating semantic information using a neural network architecture and computes replacement values for code adaptation.
Finally, Yu et al.~\cite{Yu2022} combine IR and DL techniques and propose an integrated approach, referred to as \textit{Integration} in this paper. 
\textit{Integration} first retrieves assertions using Jaccard similarity and adjusts the retrieved assertion if necessary.
Then, \textit{Integration} uses a semantic compatibility inference model to compute the ``compatibility'' of the adjusted assertion and the current focal-test. 
If the compatibility is below a specified threshold (denoted as $t$), \textit{Integration} switches to ATLAS to generate an assertion from scratch. 
The value of $t$ is determined based on the validation set.
Given that $RA_{adapt}^{NN}$ outperforms other IR-based approaches, including $IR_{ar}$ and $RA_{adapt}^{H}$, we adopt the combination of $RA_{adapt}^{NN}$ and ATLAS to explore the optimal performance of \textit{Integration}.

\section{The empirical exploration of \textit{Integration}}

The empirical study aims to investigate the application scenarios of \textit{Integration} ~\cite{Yu2022} and gain insight into its mechanisms i.e., where and why \textit{Integration} works and does not work.
To this end, three research questions are designed.

\begin{itemize}
    \item 
    \textbf{RQ1}: \textit{What are the characteristics of the dataset?}
    We first perform an in-depth analysis of the dataset. 
    In particular, we analyze the distribution of assertion lengths on the entire dataset.
    Next, for each Test-Assertion Pair $\mathrm{TAP}_{i}$ in the test set, we retrieve the assertion from the training set whose corresponding focal-test has the highest similarity to the focus-test of $\mathrm{TAP}_{i}$ and compute the edit distance between the assertion of $\mathrm{TAP}_{i}$ and the retrieved assertion.
    Such investigation not only validates the dataset's characteristics but also helps us comprehend the intrinsic association between TAPs and their similar instances, which can guide our method design (see Section~\ref{method}).
\end{itemize}



\begin{itemize}
    \item \textbf{RQ2}: \textit{Where and why does \textit{Integration} work?}
    We further explore the advantages of \textit{Integration} for assertion generation. 
    We first classify the results generated by \textit{Integration} according to the assertion generation method used. 
    Next, we analyze the assertions that are generated correctly.
\end{itemize}

\begin{itemize}
    \item \textbf{RQ3}: \textit{Where and why does \textit{Integration} fail?}
    We investigate the weaknesses of \textit{Integration}, which are critical in that understanding the application scenario of an approach can better help us apply it in practice~\cite{Liu2018}.
\end{itemize}




\begin{figure}[t]
	\graphicspath{{graphs/}}
	\centering
	\includegraphics[width=\linewidth]{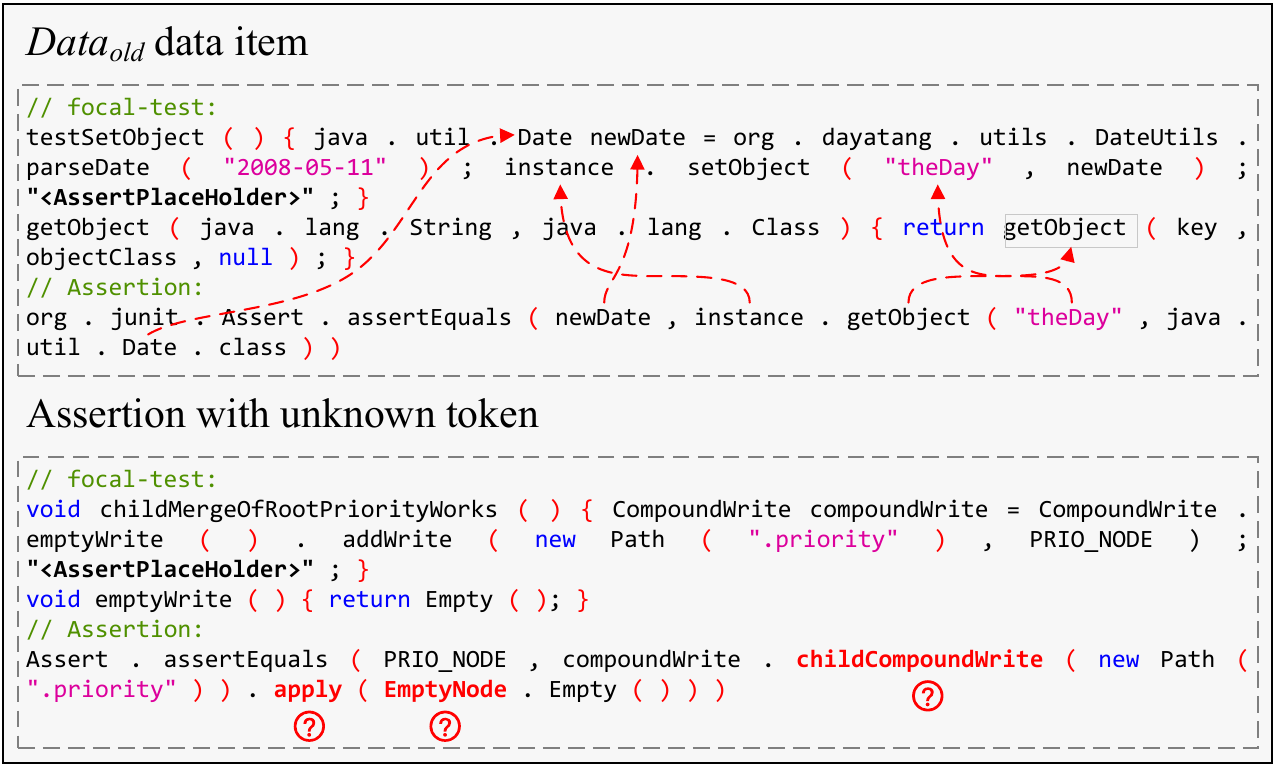}
	\caption{Assertion with known \textit{vs.} unknown tokens.}
	\label{FIG:dataset}
\end{figure}

\subsection{Dataset}
\label{dataset}

We use the two publicly available datasets~\cite{int} from Yu et al. for our experiments, namely $Data_{old}$ and $Data_{new}$, respectively.
To make our paper self-contained, we briefly describe the $Data_{old}$ and $Data_{new}$ in the following paragraphs.

\subsubsection{$Data_{old}$}

$Data_{old}$ is derived from the original dataset used by ATLAS.
Initially, $Data_{old}$ is extracted from a pool of 2.5 million test methods in GitHub, which include test prefixes and their corresponding assertion statements. 
For each test method, $Data_{old}$ includes the focal method, i.e., the production code under test.
The $Data_{old}$ is then preprocessed to exclude test methods with token lengths exceeding 1K and filter out assertions containing \textit{unknown} tokens that are not present in the focal-test and the vocabulary, following established practice in natural language processing~\cite{Merity2018, Zaremba2014}.
As an example, the bottom part of Figure~\ref{FIG:dataset} highlights the unknown tokens \texttt{childCompoundWrite}, \texttt{apply}, and \texttt{EmptyNode}.
After removing duplicates, $Data_{old}$ obtains 156,760 data items, which are further divided into training, validation, and test sets at an 8:1:1 ratio.

\subsubsection{$Data_{new}$}

The exclusion of assertions with unknown tokens can oversimplify the assertion generation problem, making $Data_{old}$ unsuitable for representing the realistic data distribution.
This, in turn, poses a significant threat to the validity of experimental conclusions.
Therefore, Yu et al.~\cite{Yu2022} constructed an expanded dataset, denoted as $Data_{new}$, by adding those cases that are excluded with unknown tokens in $Data_{old}$. 
Besides the existing data items in $Data_{old}$, $Data_{new}$ contains an extra 108,660 samples with unknown tokens to form a total of 265,420 data items, which are also divided into training, validation, and test sets in an 8:1:1 ratio.


\subsection{Study Results}

\begin{figure}[b]
	\graphicspath{{graphs/}}
	\centering
	\includegraphics[width=\linewidth]{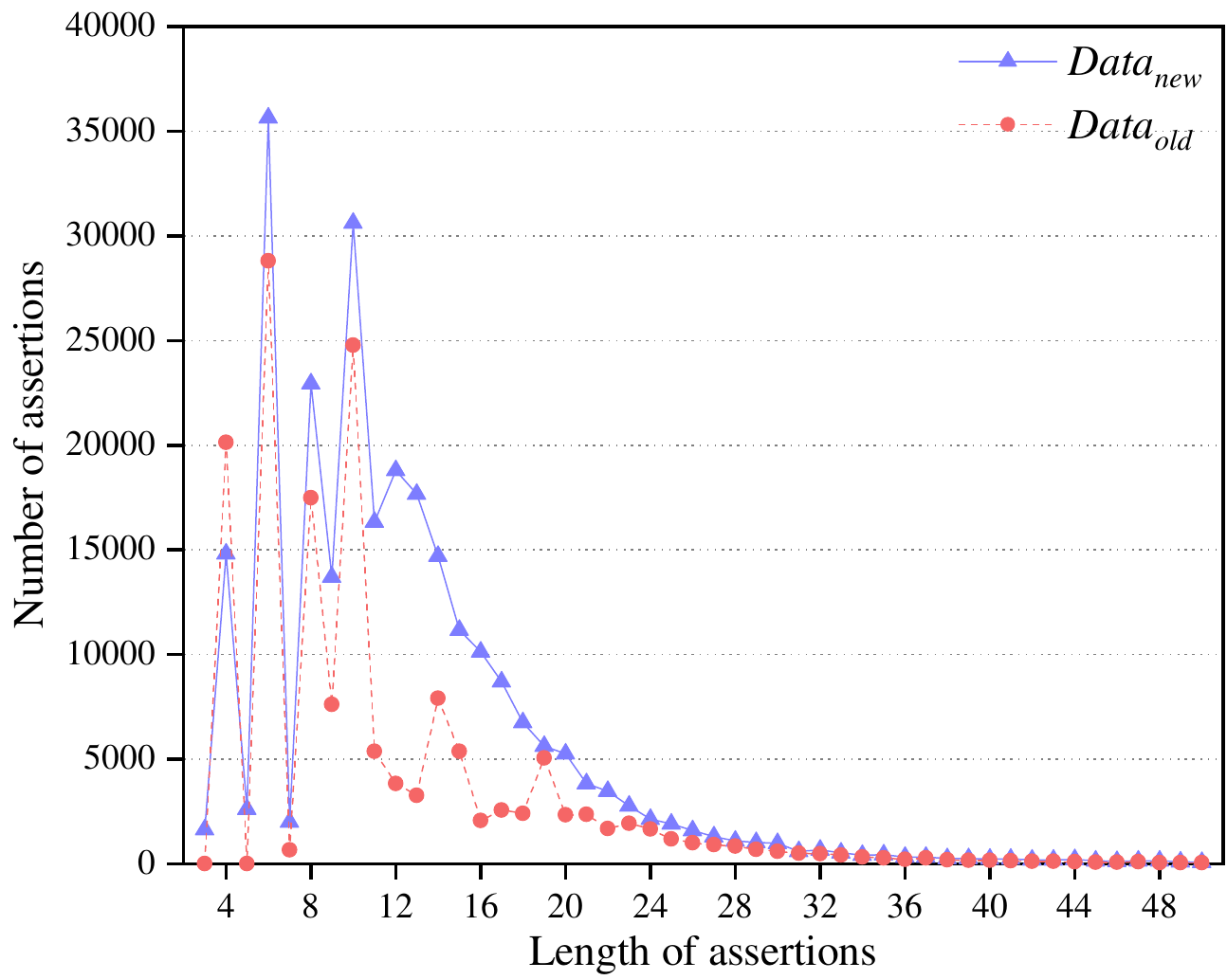}
	\caption{Length distribution of assertions of each dataset.}
	\label{FIG:assertion_length}
\end{figure}

\subsubsection{RQ1: What are the characteristics of the dataset}
Figure~\ref{FIG:assertion_length} shows the distribution of assertion length, where the X axis represents various assertion lengths and the Y axis represents the number of corresponding assertions.
Both datasets reveal a similar distribution trend: Assertions are typically less than 30 tokens, and the number of assertions decreases as the length of the assertions increases.
The average length of assertions in $Data_{new}$ is 13 tokens, while in $Data_{old}$ that is 12 tokens.
This finding suggests that in practice, testers may prefer brief assertions as long ones can cause maintenance difficulties. 


\begin{tcolorbox}[colback=cyan!5,colframe=cyan!75!black,boxsep=1mm,boxrule=0pt,top=0pt,bottom=0pt,title=Finding-1 of RQ1]
Both $Data_{new}$ and $Data_{old}$ exhibit a long-tail distribution in assertion length and edit distance.
The majority of assertion lengths are concentrated within a small number of tokens.
\end{tcolorbox}

\begin{table}[h]
\scriptsize
\setlength
\tabcolsep{1pt}
  \centering
  \caption{Edit distance (\textit{E}) between retrieved assertions and ground truth for each test set}
    \begin{tabular}{c|c|c|c|c|c|c|c|c}
    \toprule[0.7pt]
    \textbf{Dataset} & $E=0$   & $E=1$   & $E=2$   & $E=3$   & $3< E \le 5$   & $5 < E \le 10 $   & $10 < E$ & \textbf{Total}\\\hline
    $Data_{old}$ & 5,684   & 2,377       & 934        & 598         & 1,020      & 2,082         & 2,981   & 15,676 \\
    $Data_{new}$ & 10,059  & 3,059   & 1,581   & 1,031    & 1,563    & 3,050     & 6,199  &26,542\\
    \bottomrule[0.7pt]
    \end{tabular}%
  \label{tab:edit_distribution}%
\end{table}%


Table~\ref{tab:edit_distribution} further provides edit distances between test samples' ground truth (i.e., expected assertions) with retrieved assertions.
Interestingly, we can observe that a significant proportion of assertions match exactly with retrieved assertions, accounting for $37.90\% = 10,059/26,542$ and $36.26\% = 5,684/15,676$ in $Data_{new}$ and $Data_{old}$, respectively.
In addition, for $Data_{new}$ and $Data_{old}$, 65.15\% and 67.70\% of the test samples only need to modify the retrieved assertion by less than or equal to five tokens to produce a correct assertion.
Such findings provide evidence of the effectiveness of the information retrieval (IR) approach: by identifying similarities from the focal-tests, we can find almost or completely matched assertions. 


\begin{tcolorbox}[colback=cyan!5,colframe=cyan!75!black,boxsep=1mm,boxrule=0pt,top=0pt,bottom=0pt,title=Finding-2 of RQ1]
The Information Retrieval (IR) technique can successfully search nearly or completely correct assertions by identifying the similarity of focal-tests.
\end{tcolorbox}



\subsubsection{RQ2: Where and why does Integration work}

To gain insight into where and why \textit{Integration} works, we analyze its prediction results.
Given that \textit{Integration} is an integrated method that integrates ATLAS and an IR-based assertion generation method ($RA_{adapt}^{NN}$ in this paper), we categorize \textit{Integration}'s generated results based on the assertion generation method used.
Our analysis reveals that $RA_{adapt}^{NN}$ generates 21,250 ($80.06\% = 21,250/26,542$) assertions for $Data_{new}$ and 10,215 ($65.16\% = 10,215/15,676$) assertions for $Data_{old}$.
The results indicate that \textit{Integration} prefers assertions generated by the IR-based approach.


\begin{tcolorbox}[colback=cyan!5,colframe=cyan!75!black,boxsep=1mm,boxrule=0pt,top=0pt,bottom=0pt,title=Finding-1 of RQ2]
The effectiveness of \textit{Integration} is primarily dependent on the IR-based assertion generation method.
\end{tcolorbox}

\begin{table}[h]
\setlength
\tabcolsep{1pt}
  \centering
  \scriptsize
  \caption{Edit distance (\textit{E}) between retrieved assertions and assertions generated by $RA_{adapt}^{NN}$}
    \begin{tabular}{cccccccccc}
    \toprule[0.7pt]
    \textbf{Dataset} & \textbf{Prediction} & $E=0$   & $E=1$   & $E=2$   & $E=3$   & $E=4$   & $E=5$   & $E>5$   & \textbf{Total} \\
    \hline
    \multirow{2}{*}{$Data_{old}$} & \textit{Correct} & 5,435      & 486      & 341      & 108       & 73      & 35      & 40      & 6,518  \\
          & \textit{Incorrect} & 52      & 1,453      & 342      & 255      & 224      & 226       & 1,145      & 3,697  \\
    \hline
    \hline
    \multirow{2}{*}{$Data_{new}$} & \textit{Correct} & 9,839  & 437   & 235   & 67    & 32    & 8     & 22    & 10,640 \\
          & \textit{Incorrect} & 130   & 2,436  & 1,162  & 798   & 709   & 543   & 4,832  & 10,610 \\
    \bottomrule[0.7pt]
    \end{tabular}%
  \label{tab:assertion_ir}%
\end{table}%

As the IR-based approach contributes most to the \textit{Integration}, we conduct further analysis on the correct assertions generated by $RA_{adapt}^{NN}$.
Table~\ref{tab:assertion_ir} shows the edit distance between retrieved assertions and correct ones generated by $RA_{adapt}^{NN}$ for $Data_{new}$ and $Data_{old}$, accordingly.
From the table, it can be seen that the success of $RA_{adapt}^{NN}$ is mainly based on retrieving assertions that are identical to the ground truth, such as the samples with $E=0$ account for $92.47\% = 9,839/10,640$ in $Data_{new}$.
$RA_{adapt}^{NN}$ enhances the assertion generation of \textit{Integration} through adaptation operations, but is more effective for single or two token modifications than for other modification cases.
For example, with the $Data_{new}$ dataset, $RA_{adapt}^{NN}$ achieves an accuracy of $15.21\% = 437/(437+2,436)$ when only one token is altered and $7.75\% = 67/(67+798)$ when three tokens are modified.


\begin{tcolorbox}[colback=cyan!5,colframe=cyan!75!black,boxsep=1mm,boxrule=0pt,top=0pt,bottom=0pt,title=Finding-2 of RQ2]
The majority of the successful assertions produced by $RA_{adapt}^{NN}$ are exactly the same as the retrieved ones. 
In terms of adaptation operations, $RA_{adapt}^{NN}$ performs better on single token modifications than other modification cases.
\end{tcolorbox}

\subsubsection{RQ3: Where and why does Integration fail}

Previous work~\cite{Yu2022} has demonstrated that one of the bottlenecks of ATLAS's performance arises from generating assertions with long sequences. 
Hence, in RQ3, we focus on exploring where and why $RA_{adapt}^{NN}$ fails. 
We first collect the incorrect assertions generated by $RA_{adapt}^{NN}$, and calculate the edit distance between retrieved assertions and their ground truth.
As shown in Table~\ref{tab:assertion_ir}, for $Data_{new}$, there are 130 test samples 
 whose assertions match the ground truth but are still modified by $RA_{adapt}^{NN}$, and a similar phenomenon occurs in $Data_{old}$.
The performance of $RA_{adapt}^{NN}$'s adaptation strategy is limited.
For instance, when $E=1$, $RA_{adapt}^{NN}$ incorrectly generates $84.79\% = 2,436/(437+2,436)$ of assertions for $Data_{new}$ and $74.94\% = 1,453/(1,453+486)$ for $Data_{old}$.
The performance of $RA_{adapt}^{NN}$ is worse when $E>5$, with an average accuracy of only 1.91\%.

\begin{tcolorbox}[colback=cyan!5,colframe=cyan!75!black,boxsep=1mm,boxrule=0pt,top=0pt,bottom=0pt,title=Finding-1 of RQ3]
\textit{Integration} frequently incorrectly modifies assertions even when the retrieved assertion matches the ground truth exactly. 
When modifying one token to get the correct assertion, the average accuracy is only 20.14\%.
\end{tcolorbox}

\begin{table}[h]
\setlength
\tabcolsep{2pt}
  \centering
  \caption{Number of edits ($N$) made by $RA_{adapt}^{NN}$ for incorrect assertions}
    \begin{tabular}{ccccccccc}
    \toprule[0.7pt]
    \textbf{Dataset} & $N=0$   & $N=1$   & $N=2$   & $N=3$   & $N=4$   & $N=5$   & $N>5$   & \textbf{Total} \\
    \hline
    $Data_{old}$ & 3,071       & 244      & 173      & 90      & 51      & 26      & 42      & 3,697 \\
    $Data_{new}$ & 9,436  & 497   & 328   & 167   & 73    & 35    & 74    & 10,610 \\
    \bottomrule[0.7pt]
    \end{tabular}%
  \label{tab:ir_eidt}%
\end{table}%

We further count the number of edits made by $RA_{adapt}^{NN}$ for those incorrect assertions.
As reported in Table~\ref{tab:ir_eidt}, we can observe that there is a significant proportion of samples for which $RA_{adapt}^{NN}$ does not make any changes to their retrieval assertions, e.g. in $Data_{new}$, such type of sample size is 9,436, representing 88.93\% of the total number of incorrect assertions. 
Our analysis is that this may be due to the difficulty of $RA_{adapt}^{NN}$ in understanding the semantic differences between focal-tests (an example is shown in Figure~\ref{FIG:ge_example}). 
Yu et al.~\cite{Yu2022} argue that an assertion needs to be edited if it contains at least one token absent from the input focal-test. 
However, this setting ignores many instances where the necessary changes are made to adapt the context of the input focal-test, even if all the tokens of the retrieved assertions occur in the input focal-test.

\begin{tcolorbox}[colback=cyan!5,colframe=cyan!75!black,boxsep=1mm,boxrule=0pt,top=0pt,bottom=0pt,title=Finding-2 of RQ3]
\textit{Integration} fails to comprehend the semantic differences between focal-tests, resulting in many cases where retrieved assertions require modifications but are not edited appropriately.
\end{tcolorbox}

\section{Retrieve-and-Edit Assertion Generation}
\label{method}

\begin{figure*}[t]
	\graphicspath{{graphs/}}
	\centering
	\includegraphics[width=\linewidth]{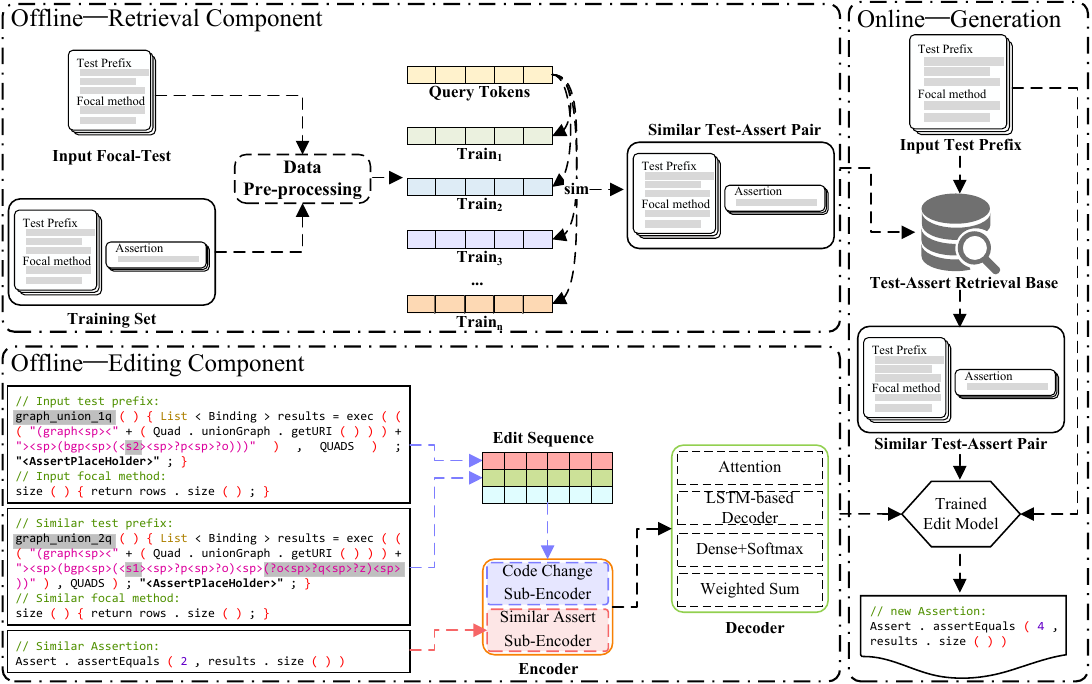}
	\caption{The overall framework of our approach.}
	\label{FIG:framework}
\end{figure*}

Drawing on insights garnered from our empirical study, we propose \textsc{EditAS}, a retrieve-and-edit approach for generating assertions.
Unlike \textit{Integration}, which only considers replacement operations to adjust retrieved assertions, \textsc{EditAS} can learn diverse assertion edit patterns from similar TAPs.
Given a focal-test as input, \textsc{EditAS} retrieves a corpus to obtain the similar TAP based on the similarity of focal-test and generates a new assertion via the retrieved TAP's assertion and compatible edit patterns.
The overall framework of \textsc{EditAS} is illustrated in Figure~\ref{FIG:framework}.
\textsc{EditAS} consists of a Retrieval component and an Edit component.

\subsection{Retrieval component}
\label{sub:retrieve}
In our approach, the Retrieve component retrieves a similar Test-Assert Pair (TAP) from a corpus given the input focal-test.
Specifically, to facilitate efficient retrieval, \textsc{EditAS} first tokenizes each focal-test in the training and test sets using javalang~\cite{javalang} and removes duplicate tokens.
Then, the Retrieve component retrieves the TAP whose focal-test has the highest Jaccard similarity coefficient with the input focal-test.
The Jaccard similarity is a text similarity measurement that considers the overlap of words between two texts, calculated using the following formula:
\begin{equation}
    J(A,B)=\frac{|A\cap B|}{|A\cup B|} 
\end{equation}
Where $A$ and $B$ are two bags of words, $|\cdot |$ denotes the number of elements in a collection.


\subsection{Edit component}
\label{sub:edit}

We employ a neural edit model to learn diverse assertion edit patterns from similar TAPs.
The edit model is designed to understand how to modify one assertion to another based on semantic differences between focal-tests.
Specifically, for a given focal-test $ft$ and its similar focal-test instance $ft'$, along with their corresponding assertions $x$ and $y$, the neural edit model aims to find a function $f$ such that $ f \left ( ft, ft', x \right ) = y$. 
In this section, we elaborate on the focal-test semantic difference representation and the neural edit model training.

\subsubsection{Focal-test semantic difference representation}

We extract and compare semantic information and modification details between focal-tests using edit sequences, according to the previous work's finding~\cite{Li2021}: different words between the two methods can reflect their semantic differences.
We follow a similar approach as in~\cite{Liu2020, Liu2023}, aligning the two tokenized focal-test sequences using a diff tool and creating an edit sequence based on the resulting alignment.
As shown in Figure~\ref{FIG:edit}, each element (named as an \textit{edit}) in an edit sequence is represented as a triple $\left \langle ft_{i}, ft_{i}{'}, a_{i} \right \rangle $, where $ft_{i}$ is a token in one focal-test and $ft_{i}{'}$ is a token in the similar focal-test, and $a_{i}$ is the edit action that transforms $ft_{i}$ to $ft_{i}{'}$.
There are four types of edit actions: \textit{insert}, \textit{delete}, \textit{equal}, or \textit{replace}. 
when $a_{i}$ is an \textit{insert} (\textit{delete}) operation, it means that $ft_{i}$ ($ft_{i}{'}$) will be an empty token $\emptyset$. 
Constructing such an edit sequence can not only preserve the information of the focal-test (i.e., $ft_{i}$ and $ft_{i}{'}$) but also highlight their fine-grained differences through $a_{i}$.

\begin{figure}[b]
	\graphicspath{{graphs/}}
	\centering
	\includegraphics[width=\linewidth]{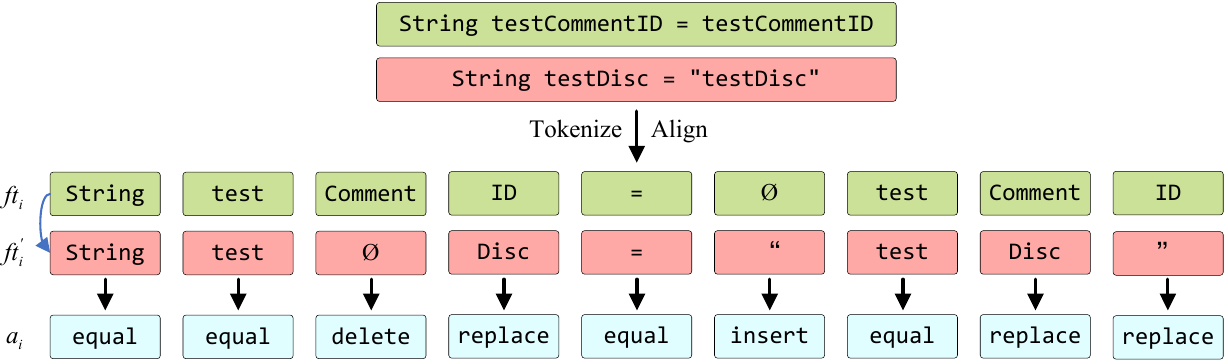}
	\caption{Converting a difference between focal-tests to an edit sequence.}
	\label{FIG:edit}
\end{figure}

\subsubsection{Neural edit model training}

Our neural edit model is fundamentally a sequence-to-sequence neural architecture.
It accepts a retrieved assertion $x=\left [ x_1, x_2, \cdots , x_{|x|} \right ]$ and an edit sequence $E = \left [ \left \langle ft_1, ft_1^{'}, a_1 \right \rangle, \dots \left \langle ft_n, ft_n^{'}, a_n \right \rangle \right ]$ as input and is designed to generate a new assertion  $ y=\left [ y_1, y_2, \cdots , y_{|y|} \right ]$.
Specifically, \textsc{EditAS} incorporates two encoders: the Edit Sequence Encoder and the Assertion Encoder, to respectively encode the edit sequence and the retrieved assertion.
An attention mechanism is then applied on the encoder side to learn the relationship between the edit sequence and the retrieved assertion. 
The decoder consists of two pointer generators that enable the model to copy tokens from both the input focal-test and the retrieved assertion concurrently during the generation process.
This approach can effectively preserve the original meaning of the retrieved assertion while integrating the assertion edit patterns reflected in the edit sequence.

\ding{182} \textbf{Encoders}.
The structures of the two encoders, i.e., the Edit Sequence Encoder and the Assertion Encoder, are nearly identical. 
They consist of a contextual embedding layer, an attention layer, and a modeling layer.

\textbf{\textit{The Contextual Embed Layer}}.
We first map the focal-test tokens, assertion tokens, and edit action tokens to embeddings. 
Considering there are only four edit actions, we randomly initialize an embedding matrix and update it during training.
To capture both syntactic and semantic information, we employ a pre-trained model, such as fastText~\cite{Grave2018}, to acquire word embeddings for each focal-test and assertion token.
We then use a bidirectional long short-term memory (Bi-LSTM)~\cite{Hochreiter1997} to process the sequence of word embeddings to access contextual information.
For Edit Sequence Encoder and each edit $E_{i}$, the three embeddings, i.e., $e_{ft_i}$, $e_{ft_{i}^{'}}$, $e_{a_i}$ are first concatenated horizontally, and then input to the Bi-LSTM, as follows:



\begin{gather*}
    e_{i}^{'} = [e_{ft_i}\oplus e_{ft_{i}^{'}}\oplus e_{a_i}] \\ 
    \overrightarrow{h}_{i}^{'} = \mathrm {LSTM}(\overrightarrow{h}_{i-1}^{'},e_{i}^{'}) ; \overleftarrow{h}_{i}^{'} = \mathrm {LSTM}(\overleftarrow{h}_{i+1}^{'},e_{i}^{'}) \\
    h_{i}^{'} = [\overrightarrow{h}_{i}^{'} \oplus \overleftarrow{h}_{i}^{'}]
\end{gather*}
where $h_{i}^{'}$ is the contextual vector of this edit and $\oplus$ is a concatenation operation.
Similarly, Assertion Encoder obtains the contextual vector $h_i$ of each assertion token $x_i$ with $x_i$'s embedding $e_{x_i}$ as input.

\textbf{\textit{The Attention Layer}}. 
This layer is responsible for linking and fusing the information of the focal-test difference and that of the retrieved assertion, capturing their relationship.
We place it on the top of the two contextual embed layers.
The attention layer takes as input the contextual vectors, i.e., $H$ and $H'$, and outputs an assertion-aware (edit-aware) feature vector for each edit (assertion token), as well as the original contextual vector for this edit (assertion token).
Formally, the edit-aware feature vector $g_i$ of assertion token $x_i$ is calculated using the dot-production attention mechanism~\cite{Luong2015} as follows:
\begin{equation}
\begin{aligned}
\label{eq1}
    &g_i = H^{'}\alpha _i;
    &\alpha_i = \mathrm{softmax}(H^{'\top}W_{\alpha}h_i)
\end{aligned}
\end{equation}
The attention weight $\alpha_i$ measures the importance of each edit is with respect to $x_i$. 
The computation of the assertion-aware feature vector $g_{i}^{'}$ of edit $E_i$ is almost same and can be expressed as follows:
\begin{gather*}
     g_{i}^{'} = H\alpha _i^{'};~~
     \alpha _i^{'} = \mathrm{softmax}(H^{\top}W_{\alpha}^{\top}h_i^{'})
\end{gather*}


\textbf{\textit{The Modeling Layer}}.
This layer uses two distinct Bi-LSTM to generate the final feature representation based on the contextual vector of each edit (assertion token) and the assertion-aware (edit-aware) feature vector, respectively. 
For example, given an assertion token $x_i$, its final representation $z_i$ is computed as follows:

\begin{gather*}
    f_{i} = [g_i\oplus h_i] \\ 
    \overrightarrow{z}_{i} = \mathrm {LSTM}(\overrightarrow{z}_{i-1},f_{i}) ; \overleftarrow{z}_{i} = \mathrm {LSTM}(\overleftarrow{z}_{i+1},f_{i}) \\
    z_{i} = [\overrightarrow{z}_{i} \oplus \overleftarrow{z}_{i}]
\end{gather*}

\ding{183} \textbf{Decoder}.
\textsc{EditAS} uses an LSTM-based decoder to generate a new assertion from the outputs of two encoders, $Z$ and $Z^{'}$. 
The final hidden states of both modeling layers are concatenated and used as the initial state for the decoder's LSTM.
During decoding step $j$, the decoder computes the hidden state $s_j$ based on the $j$-th word embedding of ground truth assertion $e_{\hat{y}_j}$, the previous hidden state $s_{j-1}$, and the previous output vector $o_{j-1}$, as follows:
\begin{equation}
    s_j = \mathrm{LSTM}(s_{j-1}, [e_{\hat{y}_j} \oplus o_{j-1}])
\end{equation}

\begin{table*}[t]
\setlength
\tabcolsep{4pt}
  \centering
  \caption{Detailed statistics of each type in $Data_{new}$ and $Data_{old}$}
    \begin{tabular}{c|c|ccccccccc}
    \toprule[0.7pt]
    \textbf{AssertType} & Total & Equals & True  & That  & NotNull & False & Null  & ArrayEquals & Same  & Other \\ \hline
    $Data_{old}$ & 15,676 & 7,866 (50\%) & 2,783 (18\%) & 1,441 (9\%) & 1,162 (7\%) & 1,006 (6\%) & 798 (5\%) & 307 (2\%) & 311 (2\%) & 2 (0\%) \\
    $Data_{new}$ & 26,542 & 12,557 (47\%) & 3,652 (14\%) & 3,532 (13\%) & 1,284 (5\%) & 1,071 (4\%) & 735 (3\%) & 362 (1\%) & 319 (1\%) & 3,030 (11\%) \\
    \bottomrule[0.7pt]
    \end{tabular}%
  \label{tab:statistics}%
\end{table*}%

We compute a context vector at each time step as the representation of the encoder's input by the dot-product attention mechanism, following Equation~\ref{eq1}.
Given there are two encoders, the decoder obtains two context vectors, i.e., $c_j$ from the retrieved assertion and $c_{j}^{'}$ from the focal-test difference.
An output vector $o_j$ is computed using $c_j$, $c_j^{'}$ and $s_j$, and the corresponding vocabulary distribution $P_{j}^{vocba}$ is obtained using a softmax layer.
\begin{gather*}
    o_j = tanh(V_c[c_j \oplus c_j^{'} \oplus s_j];~~~
    P_{j}^{vocba} = \mathrm{softmax}(V_c^{'}o_j)
\end{gather*}
where $V_c$ and $V_c^{'}$ are trainable parameters.
$P_{j}^{vocba}$ records the probability of each token being generated, of which the element with the highest probability will be the output under decoding step $j$.
Due to the similarity of focal-tests, it is reasonable to assume that certain tokens in the new assertion should also appear in the retrieved assertion, while others that are not present in the retrieved assertion should be included in the input focal-test.
Therefore, we adopt the pointer generator to copy tokens from the retrieved assertion and the input focal-test:
\begin{gather*}
    P_{j}^{ass}(y_j) = \sum_{k:x_k=y_j}\beta_{jk};~~
     P_{j}^{ft}(y_j) = \sum_{k:t_k^{'}=y_j}\beta_{jk}^{'} \\
\end{gather*}
$P_{j}^{ass}(y_j)$ and $P_{j}^{ft}(y_j)$ are the probabilities of copying $y_j$ from the retrieved assertion and the input focal-test, respectively.
$\beta_{jk}$ and $\beta_{jk}^{'}$ are the attention weights of $x_k$ and $E_k$ at time step $j$, computed based on the context vectors $c_j$ and $c_{j}^{'}$.

The conditional probability of $y_j$ at time step $j$ is then the combination of $P_{j}^{vocba}$, $P_{j}^{ass}$, and $P_{j}^{ft}$, i.e.,

\begin{displaymath}
\begin{split}
p(y_j|y_{<j},x,E)&=\gamma_{j}P_{j}^{vocba}(y_j)+(1-\gamma_{j})(\theta_jP_{j}^{ass}(y_j)+\\
&(1-\theta_j)P_{j}^{ft}(y_j))
\end{split}
\end{displaymath}
$\gamma_{j}$ and $\theta_j$ represent the probabilities of generating $y_j$ by selecting from the vocabulary and copying from the retrieved assertion, respectively. 
Both probabilities are modeled by a single-layer feed-forward neural network, which is trained jointly with the decoder.

\subsection{Generation}

Given an input focal-test $ft$, \textsc{EditAS} generates an assertion through three steps: 

\textbf{\textit{Step 1: Selecting a similar TAP.}} 
\textsc{EditAS} leverages a large-scale training dataset as the retrieval corpus. 
Then, the Retrieval component retrieves the TAP whose focal-test closely matches $ft$ from the corpus.
Further information on the retrieval process is outlined in Section~\ref{sub:retrieve}.

\textbf{\textit{Step 2: Capturing fine-grained semantic differences between focal-tests}.}
Through Step 1, \textsc{EditAS} obtains a similar focal-test to $ft$ and its corresponding assertion statements.
\textsc{EditAS} calculates the edit sequences to capture semantic differences between $ft$ and the retrieved focal-test.
The details of this part are described in Section~\ref{sub:edit}.

\textbf{\textit{Step 3: Combining retrieved assertion with semantic differences between focal-tests}.}
Finally, the trained neural edit model adjusts the retrieved assertion based on the edit sequences, thereby generating one assertion corresponding to $ft$. 
Further details on this model can be found in Section~\ref{sub:edit}.

\section{Experimental Setup}

In this section, we describe the dataset and evaluation metrics used in the experiment.
We conduct experiments to answer the following research questions:

\begin{itemize}
    \item \textbf{RQ4:} How does the proposed \textsc{EditAS} perform compared to state-of-the-art assertion generation baselines?
    \item \textbf{RQ5:} Do different similarity coefficients affect the performance of \textsc{EditAS}?
\end{itemize}

\begin{table*}[t]
  \centering
  \caption{Comparisons of our approach with each baseline}
    \begin{tabular}{c|cc|cc}
    \toprule[0.7pt]
    \multirow{2}{*}{\textbf{Approach}} & \multicolumn{2}{c|}{$Data_{old}$} & \multicolumn{2}{c}{$Data_{new}$} \\ \cline{2-5}
	         & \textbf{Accuracy} & \textbf{BLEU} & \textbf{Accuracy} & \textbf{BLEU} \\ \hline
    ATLAS               & 31.42  ($\uparrow$ 70.15\%)    & 68.51 ($\uparrow$ 17.90\%) & 21.66  ($\uparrow$ 104.80\%)       & 37.91 ($\uparrow$ 67.40\%)\\
    $IR_{ar}$           & 36.26  ($\uparrow$ 47.43\%)    & 71.48 ($\uparrow$ 13.00\%) & 37.90  ($\uparrow$ 17.04\%)       & 57.98 ($\uparrow$ 9.45\%)\\
    $RA_{adapt}^{H}$    & 40.97  ($\uparrow$ 30.49\%)    & 73.28 ($\uparrow$ 10.22\%) & 39.65  ($\uparrow$ 11.88\%)       & 59.81 ($\uparrow$ 6.10\%)\\
    $RA_{adapt}^{NN}$   & 43.63  ($\uparrow$ 22.53\%)    & 73.95 ($\uparrow$ 9.22\%) & 40.53  ($\uparrow$ 9.45\%)       & 59.81 ($\uparrow$ 6.10\%)\\
    \textit{Integration}     & 46.54  ($\uparrow$ 14.87\%)    & 78.86 ($\uparrow$ 2.42\%) & 42.20  ($\uparrow$ 5.12\%)       & 60.92 ($\uparrow$ 4.17\%) \\ \hline
    \textsc{EditAS}    & \textbf{53.46}        &  \textbf{80.77}      & \textbf{44.36}          & \textbf{63.46}  \\
    \bottomrule[0.7pt]
     \multicolumn{5}{l}{$\uparrow$ denotes performance improvement of \textsc{EditAS} against state-of-the-art baselines}
    \end{tabular}
  \label{tab:results}
\end{table*}%

\begin{table*}[t]
\setlength
\tabcolsep{3.5pt}
\scriptsize
  \centering
  \caption{Detailed statistics of our approach and each baseline for each assert type}
    \begin{tabular}{c|c|c|ccccccccc}
    \toprule
    \multirow{2}{*}{\textbf{Dataset}} & \multirow{2}{*}{\textbf{Approach}} & \multirow{2}{*}{\textbf{Total}} & \multicolumn{9}{c}{\textbf{AssertType}} \\ \cline{4-12}
	&       &       & \textbf{Equals} & \textbf{True} & \textbf{That} & \textbf{NotNull} & \textbf{False} & \textbf{Null} & \textbf{ArrayEquals} & \textbf{Same} & \textbf{Other} \\
    \midrule
    \multirow{6}{*}{$Data_{old}$} & ATLAS & 4,925 (31\%) & 2,501 (32\%) & 966 (35\%) & 248 (17\%) & 598 (51\%) & 229 (23\%) & 236 (30\%) & 100 (33\%) & 47 (15\%) & 0 (0\%) \\
          & $IR_{ar}$    & 5,684 (36\%) & 2,957 (38\%) & 1,039 (37\%) & 449 (31\%) & 439 (38\%) & 314 (31\%) & 285 (36\%) & 111 (36\%) & 89 (29\%) & \textbf{1 (50\%)} \\
          & $RA_{adapt}^{H}$   & 6,423 (41\%)   & 3,300 (42\%)   & 1,151 (41\%)      & 536 (37\%)      & 553 (48\%)      &  335 (33\%)     &  316 (40\%)      & 120 (39\%)      &  111 (36\%)     & \textbf{1 (50\%)} \\
          & $RA_{adapt}^{NN}$   & 6,839 (44\%)      & 3,509 (45\%)      & 1,225 (44\%)      & \textbf{551 (38\%)}      & 610 (52\%)      & 342 (34\%)      & 341 (43\%)       & 134 (44\%)      & 126 (41\%)      & \textbf{1 (50\%)} \\
          & \textit{Integration} & 7,295 (47\%)      & 3,714 (47\%)      & 1,333 (48\%)      & 546 (38\%)      & 724 (62\%)      & 348 (35\%)      & 352 (44\%)       & 148 (48\%)       & \textbf{129 (41\%)}      & \textbf{1 (50\%)}  \\
          & \textsc{EditAS} & \textbf{8,380 (53\%)}       & \textbf{4,131 (53\%)}      & \textbf{1,581 (57\%)}      & 526 (36\%)     & \textbf{807 (69\%)}   & \textbf{577 (57\%)}     & \textbf{469 (59\%)}      & \textbf{167 (54\%)}       & 122 (39\%)       & 0 (0\%) \\
    \hline
    \hline
    \multirow{6}{*}{$Data_{new}$} & ATLAS & 5,749 (22\%) & 2,900 (23\%) & 619 (17\%) & 537 (15\%) & 388 (30\%) & 126 (12\%) & 85 (12\%) & 47 (13\%) & 37 (12\%) & 1,010 (33\%) \\
          & $IR_{ar}$    & 10,059 (38\%) & 4,664 (37\%) & 1,436 (39\%) & 1,070 (30\%) & 600 (47\%) & 394 (37\%) & 286 (39\%) & 147 (41\%) & 113 (35\%) & 1,349 (45\%) \\
          & $RA_{adapt}^{H}$   & 10,525 (40\%)      & 4,882 (39\%)      & 1,487 (41\%)       & 1,142 (32\%)      & 651 (51\%)      & 403 (38\%)      & 297 (40\%)      & 154 (43\%)       & 121 (38\%)       & 1,388 (46\%) \\
          & $RA_{adapt}^{NN}$   & 10,758 (41\%)       & 4,988 (40\%)      & 1,526 (42\%)      & 1,161 (33\%)      & 691 (54\%)       & 401 (37\%)      & 308 (42\%)      & 162 (45\%)      & 126 (39\%)       & 1,395 (46\%)  \\
          & \textit{Integration} & 11,201 (42\%)      & 5,248 (42\%)      & 1,566 (43\%)       & 1,196 (34\%)       & 711 (55\%)       & 401 (37\%)      & 313 (43\%)      & 162 (45\%)      & 128 (40\%)      & \textbf{1,476 (49\%)}  \\
          & \textsc{EditAS} & \textbf{11,773 (44\%)}       & \textbf{5,339 (42\%)}      & \textbf{1,702 (47\%)}       & \textbf{1,304 (37\%)}       & \textbf{800 (62\%)}       & \textbf{523 (49\%)}      & \textbf{376 (51\%)}       & \textbf{172 (47\%)}       & \textbf{139 (44\%)}      & 1,418 (47\%)   \\
    \bottomrule
    \end{tabular}%
  \label{tab:type}%
\end{table*}%

\subsection{Baselines}

To answer the above-mentioned research questions, we compare \textsc{EditAS} to five baselines.
We first chose ATLAS, the first and classic neural network-based method for assertion generation.
ATLAS utilizes a sequence-to-sequence model to generate assertions from scratch.
Given that \textsc{EditAS} aims to revisit and improve retrieval-augmented deep assertion generation methods, we further adopt the three state-of-the-art (SOTA) retrieval-based methods, including $IR_{ar}$, $RA_{adapt}^{H}$, and $RA_{adapt}^{NN}$. 
Finally, we provide the performance comparison between \textsc{EditAS} and \textit{integration} which is a SOTA retrieval-augmented deep assertion generation method.
For more details please refer to Section~\ref{background}.

\subsection{Datasets}
We utilize $Data_{old}$ and $Data_{new}$ to evaluate the effectiveness of \textsc{EditAS} and baseline methods following Yu et al~\cite{Yu2022}.
Compared to $Data_{old}$, $Data_{new}$ adds the excluded cases with unknown tokens back to $Data_{old}$. 
For more details, please refer to Section~\ref{dataset}. 
Table~\ref{tab:statistics} provides detailed statistics of the test sets for the two datasets, including their distribution across different assertion types.

\subsection{Metrics}

Consistent with prior work~\cite{Watson2020, Yu2022}, the following metrics are utilized in our experiment.

\subsubsection{Accuracy}

We use accuracy to evaluate the effectiveness of assertion generation techniques. Specifically, a generated assertion is considered accurate if and only if it exactly matches the ground truth. 
Accuracy determines the percentage of samples in which the generated output matches the expected output in terms of syntax.

\subsubsection{BLEU}
In line with previous studies~\cite{Watson2020, Yu2022}, we use the muti-BLEU to measure the similarity between the generated assertion and the ground truth.
BLEU calculates the modified $n$-gram precision of a candidate sequence (i.e., the generated assertion) to the reference sequence (i.e., the ground truth), where $n$ ranges from 1 to 4.
The modified $n$-gram precision values are then averaged, and a penalty is applied for overly short sentences.

\subsection{Implementation Details}

In \textsc{EditAS}, the hyper-parameter settings are determined based on the performance on our validation set.
For the focal-test tokens, assertion tokens, and edit actions, we use 300-dimensional word embeddings.
These embeddings are obtained from a fastText model pre-trained on Common Crawl and Wikipedia~\cite{wiki}.
During training, the word embeddings are frozen.
The hidden states of the Bi-LSTMs and the LSTM in our model have dimensions of 256 and 512, respectively.
The Edit Sequence Encoder, Assertion Encoder, and Decoder are jointly trained to minimize the cross-entropy loss.
We use the Adam optimizer~\cite{Kingma2015} with a learning rate of 0.001 and clip the gradients norm by 5.
The training is done with a batch size of 8 and a dropout~\cite{Srivastava2014} rate of 0.2 for all LSTM layers.
We truncate the overlong input, where the length of the edit sequence and the focal-test is set to 512.
We stop training after 5 trials, and the model with the best (smallest) validation perplexity is selected for evaluation.
All of our approaches are built based on PyTorch framework~\cite{PyTorch}.
We conduct all the experiments on a Ubuntu 20.04 with four NVIDIA GeForce RTX 3090 GPUs, one 32-core processor, and 256GB memory.


\section{Experimental results}
\subsection{RQ4: \textsc{EditAS} \textit{vs.} Baselines}

\textit{\textbf{Overall effectiveness of \textsc{EditAS}}}.
We calculate the accuracy and BLEU scores between the assertions generated by different approaches and human-written assertions. 
The experimental results are presented in Table~\ref{tab:results}.
We notice that ATLAS performs the worst among all approaches. 
This is mainly attributed to two reasons: 
1) ATLAS, as a typical sequence-to-sequence DL model, suffers from exposure bias and gradient disappearance, leading to poor effectiveness in generating a long sequence of tokens as an assertion. 
As demonstrated in the previous work~\cite{Yu2022}, ATLAS generates less than 15 tokens with an accuracy of 46.3\%, and only 17.9\% accuracy of generating tokens with more than 15 tokens. 
2) ATLAS has a weaker ability to generate statements that contain unknown tokens, which significantly degrades its overall performance.
$IR_{ar}$ retrieves assertions from the corpus and uses them as output results, achieving better performance than ATLAS.
This indicates that the similar focal-test's assertion contains some valuable and reusable information, which also demonstrates that it is reasonable for us to use the assertion of a similar focal-test as the prototype.
$RA_{adapt}^{H}$ and $RA_{adapt}^{NN}$ further adjust the retrieved assertions to enhance the capability of the IR-based approach in generating assertions.
However, as shown in Table~\ref{tab:results}, the performance of adaptation operations of both $RA_{adapt}^{H}$ and $RA_{adapt}^{NN}$ is limited, especially for complex datasets.
For instance, $RA_{adapt}^{NN}$ can improve 20.33\% of accuracy compared to $IR_{ar}$, while for $Data_{new}$, the improvement is only 6.94\%.
A similar observation holds for $RA_{adapt}^{H}$.
\textit{Integration} combines IR and DL techniques and achieves better accuracy and BLEU scores than ATLAS and IR-based assertion generation methods.

From Table~\ref{tab:results}, \textsc{EditAS} achieves a significant performance improvement over ATLAS, with an average accuracy improvement of 87.48\% and a BLEU score improvement of 42.65\% on both datasets.
This is attributed to \textsc{EditAS} using the rich semantic information from the retrieved assertions, rather than generating assertions from scratch.
Our approach \textsc{EditAS} outperforms IR-based baseline methods and \textit{Integration} across all evaluation metrics. 
Specifically, compared to $IR_{ar}$, $RA_{adapt}^{H}$, $RA_{adapt}^{NN}$, and \textit{Integration}, \textsc{EditAS} achieves an average accuracy improvement of 32.24\%, 21.19\%, 15.99\%, and 10.00\%, respectively, which demonstrates the effectiveness of our edit module.
As compared with IR-based baselines, \textsc{EditAS} adopts the retrieved assertions as prototypes, and makes modifications by considering the semantic difference between input and similar focal-tests. 
By combining the advantages of neural networks and IR-based methods, \textsc{EditAS} achieves the best performance.

\textbf{\textit{Effectiveness on different assertion types}}.
We further compare the effectiveness of \textsc{EditAS} and baseline methods for different types of assertions. 
Each column in Table~\ref{tab:type} represents an assertion type, and each cell shows the number of correctly generated assertions and their corresponding ratios in brackets.
The results present that \textsc{EditAS} performs better than the baseline methods for almost all assertion types, especially for the standard JUnit assertion type.
For the other assertion type, \textsc{EditAS} performs marginally worse than \textit{Integration}.
This may be attributed to the fact that the training set has a large of samples of the JUnit testing framework, leading \textsc{EditAS} to learn more of the syntactic features of JUnit.
Overall, the experimental results can indicate the generality of \textsc{EditAS} in generating different types of assertions.

\begin{figure}[h]
	\graphicspath{{graphs/}}
	\centering
	\includegraphics[width=0.95\linewidth]{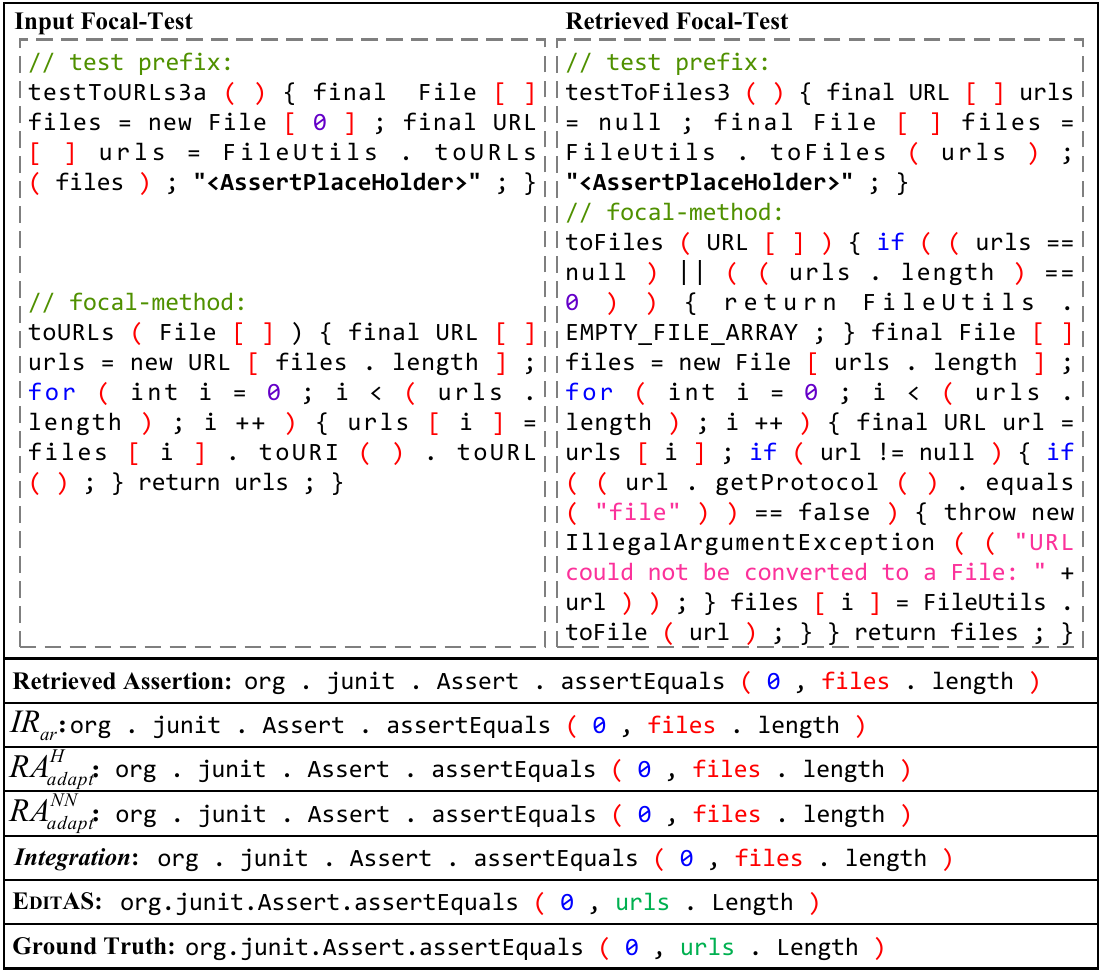}
	\caption{An example of generated assertions by our approach and baselines.}
	\label{FIG:ge_example}
\end{figure}

To better understand why \textsc{EditAS} outperforms $RA_{adapt}^{H}$ and $RA_{adapt}^{NN}$, we manually inspect the assertion generation results. 
Our analysis reveals that \textsc{EditAS} has the following advantages over the other methods: 1) \textsc{EditAS} is capable of learning and applying diverse assertion editing patterns, whereas $RA_{adapt}^{H}$ and $RA_{adapt}^{NN}$ cannot handle token addition or deletion operations.
2) $RA_{adapt}^{H}$ and $RA_{adapt}^{NN}$ only modify the retrieved assertion when it contains at least one token not present in the input focal-test. 
However, even if all tokens in the retrieved assertion appear in the input focal-test, it may still require modification due to the semantic differences between the focal-tests. In contrast, \textsc{EditAS} leverages a probabilistic model to learn common patterns of assertion edits from existing focal-tests' semantic differences.
Overall, the edit patterns learned by \textsc{EditAS} are more diverse and can cover a wider range of samples.
For example, Figure~\ref{FIG:ge_example} presents a test sample.
In this sample, the focal method of the input focal-test is responsible for transforming the file array into a URL array.
On the other hand, the retrieved focal-test performs the opposite operation. Therefore, the corresponding test prefix constructs an empty URL array to verify that the transformed files array should also be empty.
Similarly, the test prefix of the input focal-test constructs an empty files array, and its corresponding assertion aims to verify that the transformed URL array is empty.
Methods including $IR_{ar}$, $RA_{adapt}^{H}$, $RA_{adapt}^{NN}$, and \textit{Integration} encounter difficulty in understanding the semantic difference between focal-tests, as they fail to recognize the opposite operation being performed.
In addition, the tokens for the retrieved assertions all appear in the input focal-test, which does not satisfy the condition that such methods modify the assertions.
Consequently, they directly take the retrieved assertion as the expected result, leading to potential inaccuracies, while \textsc{EditAS} succeeds.



\begin{tcolorbox}[colback=cyan!5,colframe=cyan!75!black,boxsep=1mm,boxrule=0pt,top=0pt,bottom=0pt,title=Summary of RQ4]
\textsc{EditAS} significantly outperforms all baseline methods in accuracy and BLEU, with average performance improvement of 10.00\%-87.48\% and 3.30\%-42.65\% on the two datasets, respectively.
\end{tcolorbox}

\subsection{RQ5: The effectiveness with different similarity coefficients}

\textsc{EditAS} utilizes Jaccard as its default similarity coefficient. 
In this RQ, we aim to examine whether the similarity coefficients affect \textsc{EditAS}'s effectiveness. 
Specifically, we developed two additional versions of \textsc{EditAS}, utilizing two commonly used similarity coefficients, Dice~\cite{Dice1945} and Overlap~\cite{wikipedia}, respectively.
Given two sets $X$ and $Y$, Dice and Overlap calculate the similarity as follows.
\begin{gather*}
    DSC(X,Y)=|X\cap Y|/ (|X|+|Y|) \\
    Overlap(X,Y)=|X\cap Y|/min(|X|,|Y|)
\end{gather*}

Table~\ref{tab:sim} displays the accuracy of \textsc{EditAS} on both datasets using different similarity coefficients.
Our results illustrate that the similarity coefficient does have an impact on the effectiveness of \textsc{EditAS}.
Specifically, Jaccard and DICE similarity coefficients have little impact on the effectiveness of \textsc{EditAS}.
Notably, we observe that Overlap yields the poorest accuracy and BLEU scores compared to Jaccard and DICE. 
We attribute this to Overlap only accounting for the degree of overlap between the two focal-tests, ignoring their differences.

\begin{table}[h]
  \centering
  \caption{Performance of different similarity coefficients}
    \begin{tabular}{ccccc}
    \toprule[0.7pt]
    \textbf{Dataset} & \textbf{Metrics} & \textbf{Jaccard} & \textbf{Overlap} & \textbf{DICE} \\
    \hline
    \multirow{2}{*}{$Data_{old}$} & Accuracy & 53.46       & 46.13      & 53.46 \\
          & BLEU  & 80.77      & 75.42      & 80.77 \\
    \hline
    \hline
    \multirow{2}{*}{$Data_{new}$} & Accuracy & 44.36      & 38.22      & 44.24 \\
          & BLEU  & 63.46      & 56.37      & 63.59 \\
    \bottomrule[0.7pt]
    \end{tabular}%
  \label{tab:sim}%
\end{table}%

\begin{tcolorbox}[colback=cyan!5,colframe=cyan!75!black,boxsep=1mm,boxrule=0pt,top=0pt,bottom=0pt,title=Summary of RQ5]
The similarity coefficient does have an impact on the effectiveness of \textsc{EditAS}.
Improving the Retrieval component's capability can lead to better assertion generation performance of \textsc{EditAS}.
\end{tcolorbox}

\section{Threats to validity}
There are three main threats to the validity of our approach. 
Firstly, following previous works~\cite{Yu2022, Watson2020}, we only conducted experiments on two Java datasets.
Although Java may not be representative of all programming languages, the datasets used in our experiments are large enough and provide sufficient safety to demonstrate the effectiveness of \textsc{EditAS}.
Furthermore, \textsc{EditAS} employs exclusively language-agnostic features, i.e., edit sequences, and can be applied to other programming languages.
Secondly, the Retrieve component retrieves a similar focal-test based on lexical similarity. 
This may result in the retrieved focal-test and input focal-test being similar only at the lexical level while exhibiting different assertions. 
To address this threat, we use a large-scale Java dataset (247M) to increase the scale and diversity of the retrieval corpus. 
We also propose an Edit component, which modifies the prototype by considering the semantic differences between the input focal-test and the retrieved focal-test, to alleviate this threat.
The final major threat comes from the lack of comparing \textsc{EditAS} with other existing assertion generation methods in terms of generating compilable assertions.
While compilability can be used as another indicator of assertion quality, automated construction has been a challenging task that is highly dependent on external/internal settings/resources~\cite{Hassan2018, Lou2020}.
Hence, automatically compiling large-scale assertions remains an obstacle.
To ensure scalability, we do not use compilability and bug detection as metrics.

\section{Conclusion and Future Works}

In this paper, we reaffirm the previous finding that Information Retrieval (IR) can always find a ``almost correct'' assertion that is very similar to the expected one, and emphasize the shortcomings of previous approaches in modifying retrieved assertions.
To alleviate these problems, we propose a novel retrieve-and-edit approach named \textsc{EditAS} for assertion generation. 
\textsc{EditAS} contains two components. 
A Retrieve component for retrieving the similar focal-test that consists of a test method without any assertion and its focal method (i.e., the method under test).
An Edit component treats the assertion of a similar focal-test as a prototype and combines the prototype and assertion edit pattern reflected by semantic differences between the input focal-test and similar focal-test to generate a target assertion.
We conducted extensive experiments on two large-scale Java datasets. 
The experimental results show that \textsc{EditAS} substantially outperforms the state-of-the-art baselines. 
Our work shows that for the assertion generation task, retrieving similar assertions and learning to modify retrieved assertions by applying a set of editing operations can achieve satisfactory performance.
In the future, we plan to explore additional techniques to enhance \textsc{EditAS} further.
One avenue we will explore is integrating contextual information or leveraging code language models to handle code changes more effectively.
Additionally, we aim to extend our approach to support other programming languages, such as Python, to enhance the generalizability and applicability of our method.
Our source code and experimental data are available at \href{https://github.com/swf-cqu/EditAS}{https://github.com/swf-cqu/EditAS}.

\section{Acknowledgements}
We appreciate the insightful insights provided by anonymous reviewers to improve the quality of the paper.
This work was supported in part by the National Key Research and Development Project (No. 2021YFB1714200), the Fundamental Research Funds for the CentralUniversities (No. 2022CDJDX-005), the Chongqing Technology Innovation and Application Development Project (No. CSTB2022TIAD-STX0007 and No. CSTB2022TIAD-KPX0067), the National Natural Science Foundation of China (No. 62002034) and the Natural Science Foundation of Chongqing (No. cstc2021jcyj-msxmX0538).

\bibliographystyle{IEEEtran.bst}
\balance

\bibliography{ref}

\end{document}